\documentclass{PoS}

\let\OLDthebibliography\thebibliography
\renewcommand\thebibliography[1]{
  \OLDthebibliography{#1}
  \setlength{\parskip}{0pt}
  \setlength{\itemsep}{0pt plus 0.3ex}
  \footnotesize
}

\usepackage{siunitx}
\usepackage{wrapfig}
\usepackage{epsfig,epstopdf}
\usepackage{graphicx, color}
\usepackage{subcaption}
\usepackage[font={footnotesize,it}]{caption}
\usepackage[colorinlistoftodos,prependcaption,textsize=tiny]{todonotes}

\title{Matter-wave Atomic Gradiometer Interferometric Sensor (MAGIS-100) at Fermilab}

\ShortTitle{(MAGIS-100) at Fermilab}

\author{\speaker{Jon Coleman}\thanks{On Behalf of the MAGIS-100 Collaboration. }\\
        University of Liverpool\\
        E-mail: \email{coleman@liverpool.ac.uk}}

\abstract{MAGIS-100 is a  next-generation instrument that uses light-pulse atom interferometry to search for physics beyond the standard model, to be built and installed at Fermilab. We propose to search for dark matter and new forces, and to test quantum mechanics at new distance scales. The detector will use the existing \SI{100}{m} vertical NuMI access shaft to make it the world's longest baseline atom interferometer. To maximize the sensitivity of the experiment, we will use the latest advances in atomic clock technologies. The experiment will be a significant step towards developing a \SI{1000}{m} baseline detector, with sufficient sensitivity to detect gravitational waves in the `mid-band' from \SI{0.1}{Hz} - \SI{10}{Hz}, between the Advanced LIGO and LISA experiments. Here we describe an overview of the experiment and its physics reach.}

\FullConference{The 39th International Conference on High Energy Physics (ICHEP2018)\\
		4-11 July, 2018\\
		Seoul, Korea}

\begin{document}
\vspace{-5mm}
\section{Introduction}

\vspace{-1mm}

MAGIS-100 is a new experiment at Fermilab to search for physics beyond the standard model using atom interferometry~\cite{PhysRevLett.67.181}. The aim is to look for dark matter and new forces by exploiting atom interferometry's unique sensitivity to external potentials at the level of \SI{e-20}{eV}~\cite{arvanitaki2016search,Graham:2015ifn,Graham:2017ivz}. The scientific goals will be accomplished by connecting two \SI{50}{m} interferometers across a vertical baseline \SI{100}{m} and by achieving unprecedented macroscopic quantum superpositions~\cite{kovachy2015quantum}. The output, stored as a relative phase in each interferometer, is compared across the baseline, enabling the removal of common-mode background noise. Due to the complexity of scaling these devices to the \SI{100}{m} scale and beyond, such an experiment has not yet been constructed.

\begin{figure}
\centering
\begin{subfigure}{.74\textwidth}
  \centering
  \includegraphics[width=0.75\linewidth]{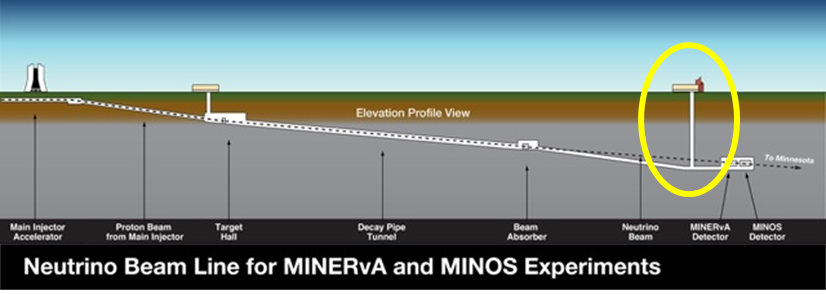}
\end{subfigure}
\hspace{1pt}
\begin{subfigure}{.2\textwidth}
  \centering
  \includegraphics[width=.75\linewidth]{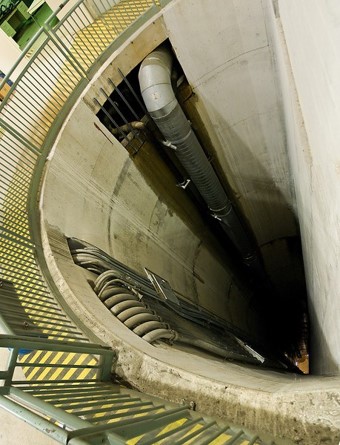}
\end{subfigure}
\caption{(Left) Proposed site for MAGIS-100. Elevation profile view of existing NuMI (Neutrino Main Injector) tunnel at Fermilab.  MAGIS-100 will be located in the  \SI{88}{m} access shaft, denoted by the yellow circle. (Right) Photograph of the top of the shaft - the detector is to be mounted vertically along the wall.}
\label{fig:numi}
\end{figure}

To realize this experiment, we will use the existing NuMI access shaft at Fermilab, shown in Fig.~\ref{fig:numi}. The site is currently being measured to characterize the environment, including vibration, temperature variation, and magnetic fields.

The experiment is a milestone towards the construction of a future \SI{1}{km} detector sensitive to gravitational waves~\cite{Graham:2012sy} in the `mid-band' frequency range (\SI{0.1}{Hz} - \SI{10}{Hz}) between the Advanced LIGO and proposed LISA experiments.

\vspace{-5mm}

\section{Physics}

\subsection{Dark Matter}
The energy budget of the universe is dominated by 73\% dark energy  and 23\% dark matter, the nature of which remains an enigma, and has so far evaded direct detection. Given the null results from the present generation of dark matter experiments, it is important to broaden the search to cover a wider mass range. Well motivated theories indicate that the mass range from \SI{e-22}{eV} to \SI{e-3}{eV} is particularly interesting. Potential candidates within this range include the QCD axion, axion-like-particles, and the relaxion. Dark matter in this mass range can be described as a field that oscillates at a frequency determined by the mass of the dark matter particle, resulting in time dependent effects. As the field oscillates, the properties of the atoms coupled to the field (such as the quantum energy level and spin) also change, leading to a measurable signal set by the the mass of the dark matter, enabling  multiple high precision searches for ultra-weak effects.

\begin{figure}
\centering
\begin{subfigure}{.45\textwidth}
  \centering
 \includegraphics[width=0.8\textwidth]{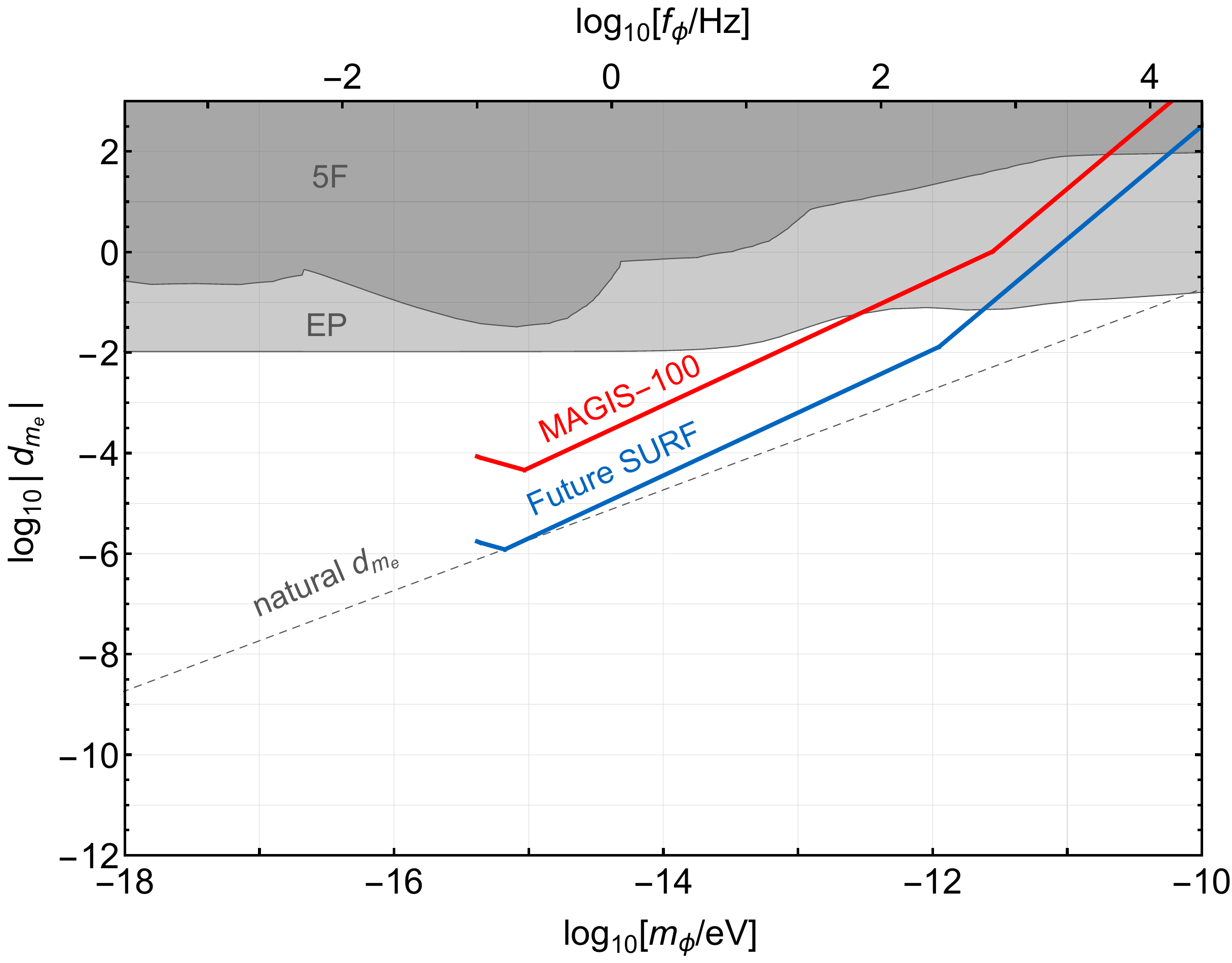}
\end{subfigure}
\hspace{2pt}
\begin{subfigure}{.45\textwidth}
  \centering
  \includegraphics[width=0.8\textwidth]{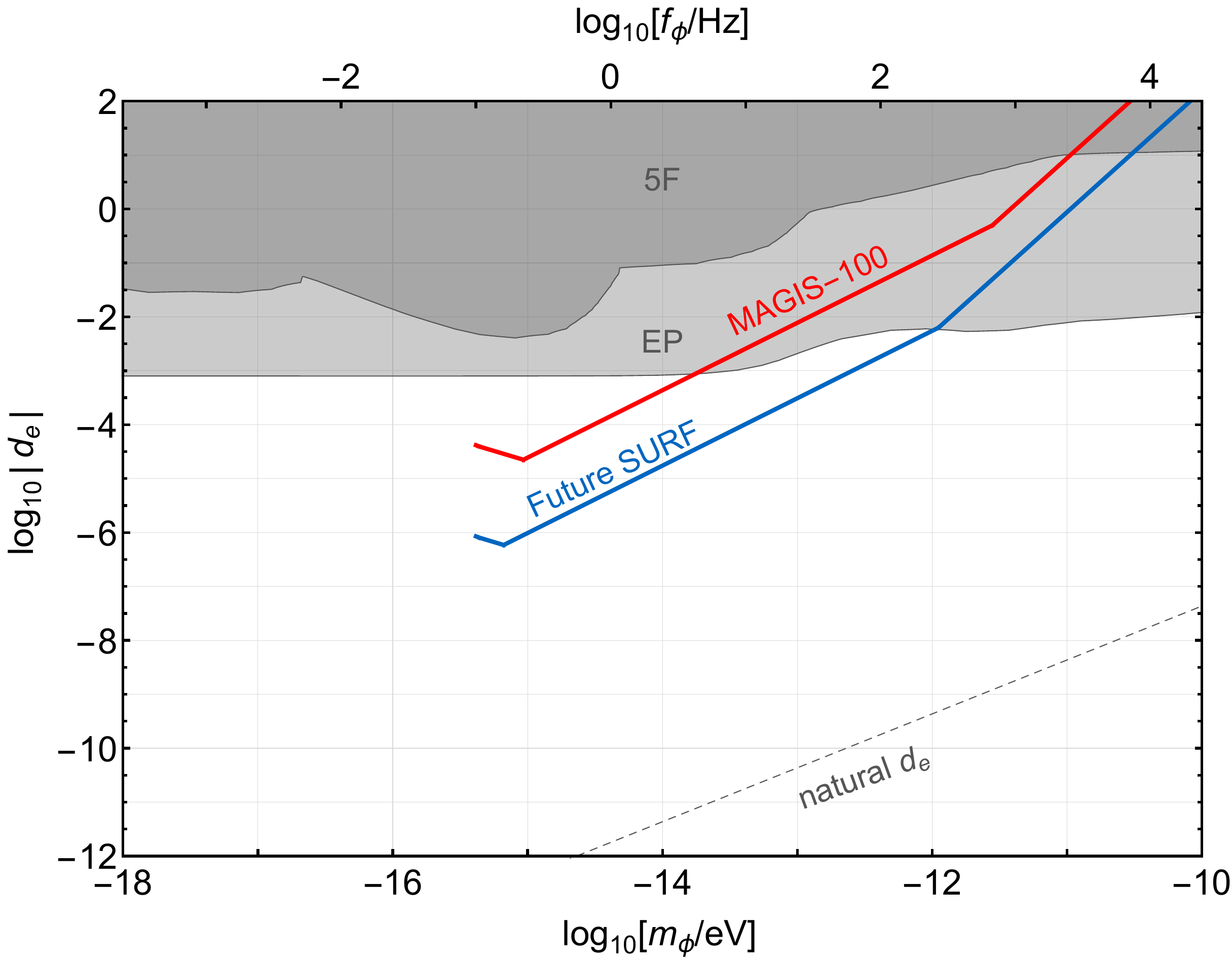}
\end{subfigure}
\caption{
Sensitivity to an ultralight dark matter field coupling to the electron mass with strength $d_{m_e}$ (left), and sensitivity to dark matter via coupling to the fine structure constant with  strength $d_e$ (right), shown as a function of the mass of the scalar field $m_\phi$ (or alternatively the frequency of the field - top scale).  The red curve, which shows the reach for an exposure of \num{e15} dropped atoms, assumes a shot-noise limited phase resolution and corresponds to about 1 year of data taking at MAGIS-100. The gray bands show bounds derived from equivalence principle (EP) and fifth force (5F) tests. The blue curve is the projected sensitivity of a future kilometer-scale detector.}
\label{fig:dm-sensitivity}
\vspace{-3mm}
\end{figure}

First, dark matter that affects fundamental constants, e.g. the electron mass or the fine structure constant, will change the energy levels of the quantum states used in an interferometer.  In MAGIS-100, this effect will be detected by  simultaneously comparing the outputs of the two interferometers~\cite{arvanitaki2016search}.  The sensitivity to several such dark matter candidates is shown in Fig.~\ref{fig:dm-sensitivity}. In the mass range $\SI{e-14}{eV} < m_\phi < \SI{e-15}{eV}$, MAGIS-100 will improve on existing bounds by at least two orders of magnitude. 

\begin{wrapfigure}{L}{0.48\textwidth}
\begin{center}
\vspace{-5 mm}
\includegraphics[width=0.45\textwidth]{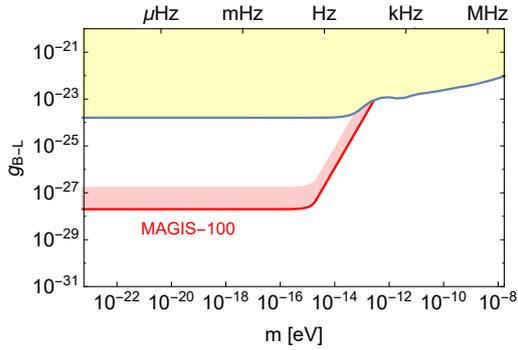}
\caption{Sensitivity to a B-L coupled new force, with  $10^{-16}g/\sqrt{\text{Hz}}$ sensitivity (assumes $50~\text{m}$ launch, $1000~\hbar k$ atom optics, $10^8~\text{atoms/s}$ flux, shot noise limited). Shaded red band shows estimated uncertainty in projected sensitivity.  Yellow band indicates existing bounds.  Sensitivities of this method to other dark matter candidates are shown in \cite{Graham:2015ifn}. }
\label{fig:BL-DM-sensitivity}
\vspace{-13mm}
\end{center}
\end{wrapfigure}

Second, dark matter can be searched for by comparing the accelerometer signals from two simultaneous quantum interferometers run with different isotopes \cite{Graham:2015ifn}.  This  requires running a dual-species atom interferometer, \cite{PhysRevLett.113.023005,PhysRevA.88.043615,1367-2630-16-7-073035,overstreet2018effective}. The potential sensitivity to one such candidate, a B-L coupled new vector boson, is shown in Fig.~\ref{fig:BL-DM-sensitivity}. Note that, compared to existing bounds,  MAGIS-100  {\parfillskip0pt\par} 

\noindent will  improve the sensitivity to any such dark matter particles with mass (frequency) below approximately \SI{e-15}{eV} (\SI{0.1}{Hz}) by about four orders of magnitude.  

Third, dark matter that causes precession of nuclear spins, such as pseudo-scalar dark matter, can be searched for by comparing simultaneous, co-located interferometers using  atoms in quantum states with differing nuclear spins.  See \cite{Graham:2017ivz} for a discussion and potential sensitivities.
\vspace{-1mm}

\subsection{Gravitational Waves}

\begin{wrapfigure}{R}{0.45\textwidth}
\footnotesize
\begin{center}
\vspace{-7mm}
\includegraphics[width=0.45\textwidth]{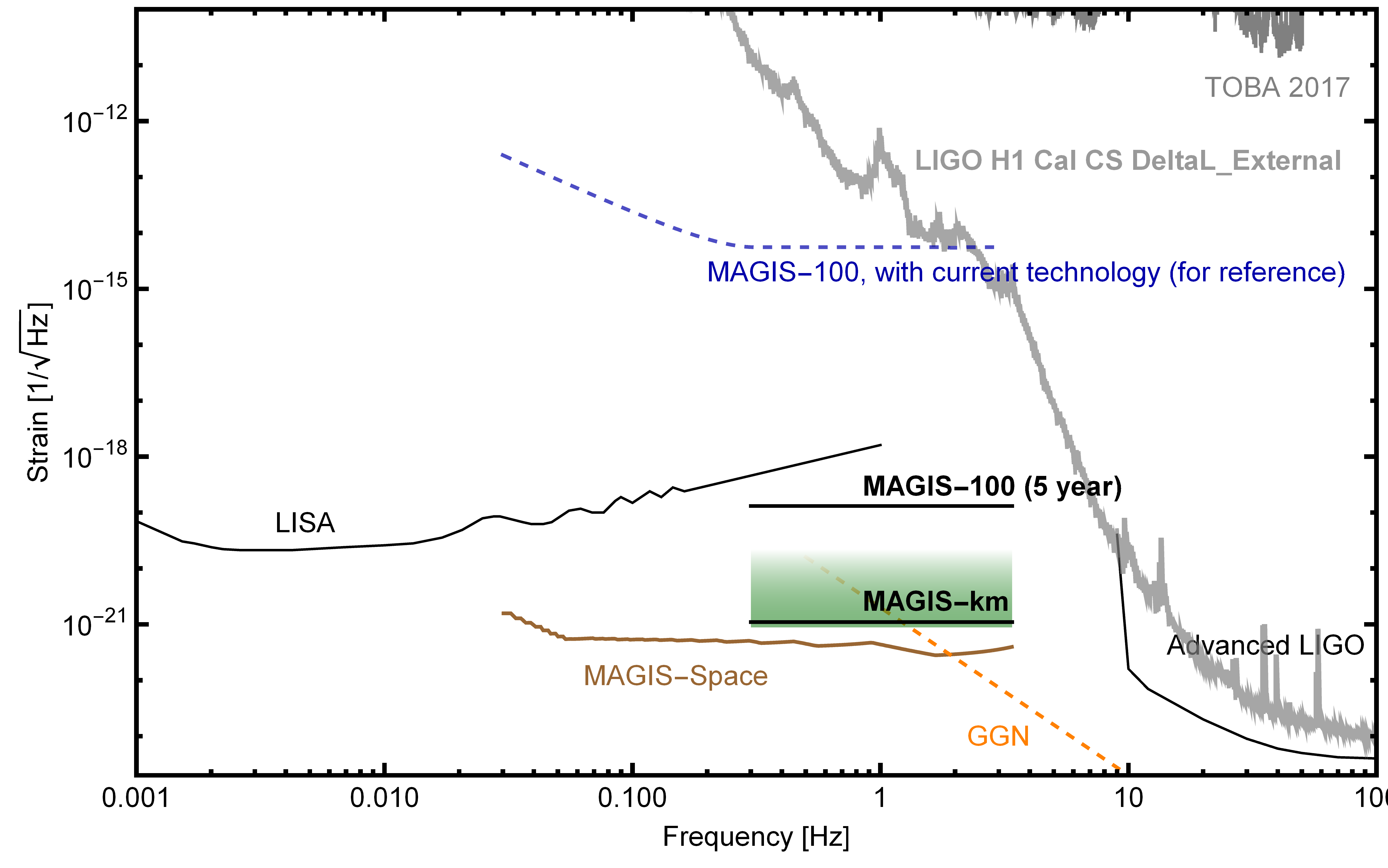}
\end{center}
\vspace{-6mm}
\caption{Gravitational wave sensitivity, characterized by the strain sensitivity,  shown as a function of frequency. The blue-dash curve shows the strain sensitivities using current parameters ($100~\hbar k$, $\delta\phi=10^{-3}~\text{rad}/\sqrt{\text{Hz}}$).  The curve labelled MAGIS-100 5 year. shows what is possible after sensor research and development. Also shown is the estimated sensitivity of a future km-scale experiment.  Existing limits from TOBA \cite{PhysRevD.95.082004}, and Advanced LIGO and LISA sensitivity curves are shown for reference. A preliminary estimate of Newtonian gravitational gradient noise (GGN), expected to limit future terrestrial detectors at low frequencies, is shown in orange. 
}
\label{fig:gw-sensitivity-100}
\footnotesize
\vspace{-10mm}
\end{wrapfigure}

The extension of MAGIS-100 schema to a  \SI{1000}{m} device, enhances the sensitivity to new physics and gravitational waves, see Fig.~\ref{fig:gw-sensitivity-100}. The SURF laboratory in South Dakota provides a candidate location for such a \mbox{MAGIS-}1000 experiment. The construction of \SI{1000}{m} interferometer would access the highest energy scales in the very early universe, being above the white dwarf ``confusion noise", but low enough in frequency to see cosmological sources. The mid-band is an excellent place to search for gravitational waves from inflation and reheating, and  models of axion inflation also give signals large enough to be detected. Furthermore, gravitational wave signals in this band may be produced by phase transitions in the early universe at scales 
{\parfillskip0pt\par}

\noindent above the weak scale and by networks of cosmic strings, providing sensitivity to physics of the early universe.
\vspace{-2mm}
\section{Summary}
\vspace{-2mm}
A next-generation atom interferometer is being developed at Fermilab. Design, construction, and integration of the main components of MAGIS-100 is expected to take two to three years. The initial science program will then take an additional three years. It will provide an increase of two orders of magnitude in sensitivity of a dark matter field in the mass region \SI{e-15}{eV} and four orders of magnitude increase in sensitivity to B-L coupled new forces over eight orders of magnitude in mass. The experiment will also yield sensitivity to pseudoscalar dark matter. MAGIS-100 is the precursor to a proposed \SI{1000}{m} experiment which will provide leading sensitivity to gravitational waves in the mid-band frequency range. The project will advance the frontier of quantum sensor technologies and the physics that can be pursued with these technologies, including quantum tests on record-setting macroscopic scales.

\vspace{-2mm}
\bibliographystyle{unsrt}
\bibliography{bibliography}

\begin{thebibliography}{10}

\bibitem{PhysRevLett.67.181}
M.~Kasevich and S.~Chu.
\newblock { Phys. Rev. Lett.} {\bf 67}, 181 (1991).

\bibitem{arvanitaki2016search}
A.~Arvanitaki, P.~Graham, J.~Hogan, S.~Rajendran, and K.~Van~Tilburg.
\newblock { Phys. Rev. D} {\bf 97}, 075020 (2018).

\bibitem{Graham:2015ifn}
P.~Graham, D.~Kaplan, J.~Mardon, S.~Rajendran, and W.~A. Terrano.
\newblock { Phys. Rev. D} {\bf 93}, 075029 (2016).

\bibitem{Graham:2017ivz}
P.~Graham, D.~Kaplan, J.~Mardon, S.~Rajendran, W.~A. Terrano, L.~Trahms, and
  T.~Wilkason.
\newblock { Phys. Rev. D} {\bf 97}, 055006 (2018).

\bibitem{kovachy2015quantum}
T.~Kovachy, P.~Asenbaum, C.~Overstreet, C.~Donnelly, S.~Dickerson,
  A.~Sugarbaker, J.~Hogan, and M.~Kasevich.
\newblock {Nature} {\bf 528}, 530 (2015).

\bibitem{Graham:2012sy}
P.~Graham, J.~Hogan, M.~Kasevich, and S.~Rajendran.
\newblock { Phys. Rev. Lett.} {\bf 110}, 171102 (2013).

\bibitem{PhysRevLett.113.023005}
M.~G. Tarallo, T.~Mazzoni, N.~Poli, D.~V. Sutyrin, X.~Zhang, and G.~M. Tino.
\newblock { Phys. Rev. Lett.} {\bf 113}, 023005 (2014).

\bibitem{PhysRevA.88.043615}
A.~Bonnin, N.~Zahzam, Y.~Bidel, and A.~Bresson.
\newblock { Phys. Rev. A} {\bf 88}, 043615 (2013).

\bibitem{1367-2630-16-7-073035}
C.~C.~N. Kuhn, G.~D. McDonald, K.~S. Hardman, S.~Bennetts, P.~J. Everitt, P.~A.
  Altin, J.~E. Debs, J.~D. Close, and N.~P. Robins.
\newblock {New Journal of Physics} {\bf 16}, 073035 (2014).

\bibitem{overstreet2018effective}
C.~Overstreet, P.~Asenbaum, T.~Kovachy, R.~Notermans, J.~Hogan, and
  M.~Kasevich.
\newblock {Phys. Rev. Lett} {\bf 120}, 183604 (2018).

\bibitem{PhysRevD.95.082004}
A.~Shoda, Y.~Kuwahara, M.~Ando, K.~Eda, K.~Tejima, Y.~Aso, and Y.~Itoh.
\newblock {Phys. Rev. D} {\bf 95}, 082004 (2017).

\bibitem{PAC}
P.~{Adamson,}~{\it et al}.
\newblock {Letter of {I}ntent: {Fermilab P-1101, Matter-wave Atomic Gradiometer
  Interferometric Sensor ({MAGIS-100} Collaboration), (2018)}}.

\end{thebibliography}
\footnotesize

\noindent{\bf Acknowledgements} This presentation and proceedings are based on the MAGIS-100 Collaboration LoI, presented to the Fermilab PAC~\cite{PAC}.  Fermi National
Accelerator Laboratory (Fermilab), is a U.S. Department of
Energy, Office of Science, HEP User Facility. Fermilab is
managed by Fermi Research Alliance, LLC (FRA), acting
under Contract No. DE-AC02-07CH11359. The work was supported in part by the Royal Society, UK, and the Universities Research Alliance (URA).
\end{document}